\documentclass[letter]{aa} 

\usepackage{epsfig}     
\usepackage{graphicx,color}     
\usepackage{amssymb}            
\usepackage{url}                
\usepackage{amsmath}            
\usepackage{rotating}                   
\usepackage{float}                      
\usepackage{textcomp}           
\usepackage{epstopdf}
\usepackage{dcolumn}
\usepackage{times}
\usepackage{tabularx}
\usepackage{hyperref}
\hypersetup{
    colorlinks,
    citecolor=blue,
    filecolor=blue,
    linkcolor=blue,
    urlcolor=blue,
    menucolor=black}
\usepackage{soul} 
\usepackage[english]{babel}
\usepackage{booktabs}
\usepackage{gensymb}
\usepackage{subfig}

\begin{document}


\title{What drives decayless kink oscillations in active region coronal loops on the Sun?}

\author{S.~Mandal\inst{1},
L.~P.~Chitta\inst{1}, 
P.~Antolin\inst{2},
H.~Peter\inst{1},
S.~K.~Solanki\inst{1,3},
F.~Auch\`{e}re\inst{4}, 
D.~Berghmans\inst{5}, 
A.~N.~Zhukov\inst{5,6},
L.~Teriaca\inst{1}, 
R.~Aznar Cuadrado\inst{1}, 
U.~Sch\"{u}hle\inst{1},
S.~Parenti\inst{4}, 
\'{E}.~Buchlin\inst{4},
L.~Harra\inst{7,8}, 
C.~Verbeeck\inst{5},
E.~Kraaikamp\inst{5},
D.~M.~Long\inst{9},
L.~Rodriguez\inst{5},
G.~Pelouze\inst{4},
C.~Schwanitz\inst{7,8},
K.~Barczynski\inst{7,8},
\and
P.~J.~Smith\inst{9}
}

\institute{Max Planck Institute for Solar System Research, Justus-von-Liebig-Weg 3, 37077, G{\"o}ttingen, Germany \\
\email{smandal.solar@gmail.com}
\and
Department of Mathematics, Physics and Electrical Engineering, Northumbria University, Newcastle upon Tyne, NE1 8ST, UK
\and
School of Space Research, Kyung Hee University, Yongin, Gyeonggi 446-701, Republic of Korea
\and
Institut d'Astrophysique Spatiale, CNRS, Univ. Paris-Sud, Universit\'{e} Paris-Saclay, B\^{a}t. 121, 91405 Orsay, France
\and
Royal Observatory of Belgium, Ringlaan -3- Av. Circulaire, 1180 Brussels, Belgium
\and
Skobeltsyn Institute of Nuclear Physics, Moscow State University, 119992 Moscow, Russia
\and
Physikalisch-Meteorologisches Observatorium Davos, World Radiation Center, 7260 Davos Dorf, Switzerland 
\and
ETH-Z\"{u}rich, Wolfgang-Pauli-Str. 27, 8093 Z\"{u}rich, Switzerland
\and
UCL-Mullard Space Science Laboratory, Holmbury St. Mary, Dorking, Surrey RH5 6NT, UK
}

\abstract{
We study here the phenomena of decayless kink oscillations in a system of active region (AR) coronal loops. Using high resolution observations from two different instruments, namely the Extreme Ultraviolet Imager (EUI) on board Solar Orbiter and the Atmospheric Imaging Assembly (AIA) on board the Solar Dynamics Observatory, we follow these AR loops for an hour each on three consecutive days. Our results show significantly more resolved decayless waves in the higher-resolution EUI data compared with the AIA data. Furthermore, the same system of loops exhibits many of these decayless oscillations on Day-2, while on Day-3, we detect very few oscillations and  on Day-1, we find none at all. Analysis of photospheric magnetic field data reveals that at most times, these loops were rooted in sunspots, where supergranular flows are generally absent. This suggests that supergranular flows, which are often invoked as drivers of decayless waves, are not necessarily driving such oscillations in our observations. Similarly, our findings also cast doubt on other possible drivers of these waves, such as a transient driver or mode conversion of longitudinal waves near the loop footpoints. In conclusion, through our analysis we find that none of the commonly suspected sources proposed to drive decayless oscillations in active region loops seems to be operating in this event and hence, the search for that elusive wave driver needs to continue. 
}

   \titlerunning{What drives decayless kink oscillations in active region coronal loops on the Sun?}
   \authorrunning{Sudip Mandal et al.}
   \maketitle

\section{Introduction}
Decayless kink oscillations (or simply, decayless oscillations) are generally observed to be small amplitude kink oscillations exhibited by coronal loops. These oscillations are referred to as decayless because they show no significant decay in their amplitudes over multiple wave periods \citep{2012ApJ...759..144T,2013A&A...560A.107A}. More generally, decayless oscillations are observed in the absence of any nearby transients (exceptions being the oscillations reported in \citealp{2012ApJ...751L..27W} and \citealp{2021A&A...652L...3M}).
These properties are in stark contrast to the rapidly decaying kink oscillations associated with flares and/or eruptions \citep{1999Sci...285..862N, 1999ApJ...520..880A}. To persist over multiple wave periods, these decayless oscillations must overcome damping in the corona (e.g. via resonant absorption or wave dissipation). However, since their discovery, the reason for such persistent presence remains unknown.

Over the years, a number of theoretical and numerical studies have been performed to investigate the origin of these decayless oscillations. While \citet{2013A&A...552A..57N} suggested that a harmonic or a quasi-harmonic driver at the loop footpoints or in the lower solar atmosphere, could generate the observed oscillations, \citet{2015A&A...583A.136A} argued in favour of a random footpoint driving. In any case, both these models fail to capture many observed properties of these oscillations. \citet{2016A&A...591L...5N} proposed a self-oscillation model in which the loop footpoints are driven by supergranular motions, analogous to the sliding of a bow on a violin. In such a scenario, the oscillation period is set by the system itself and not by the driver. In simple terms, a non-periodic driver can produce periodic oscillations when the oscillating system is in a self-oscillation mode \citep{2013PhR...525..167J}. This idea was later adopted in a 3-D numerical simulation by \citet{2020ApJ...897L..35K} who managed to produce basic observational features such as the decayless behaviour of oscillations, their periods and amplitudes, through a proof-of-concept model. 
Another way a loop could reach a self-oscillating state is via Alfv\'enic vortex shedding, a concept first put forward by \citet{2009A&A...502..661N}.  3-D numerical modelling based on this idea was performed by \citet{2021ApJ...908L...7K}. In this case, a strong background flow generates vortices near the loop boundary that are then responsible for driving transverse loop displacements, akin to decayless oscillations. Similar to earlier models, a steady flow generates periodic oscillations, a key feature of the self oscillation process. The other idea related to such decayless wave generation is through a combination of resonant absorption and Kelvin-Helmholtz instabilities (KHI) induced by the kink waves near the loop boundaries, which lead to an apparent decayless oscillation at low spatial resolution  \citep{2016ApJ...830L..22A}. However, no consensus has been reached yet on how these decayless oscillations actually originate \citep{2021SSRv..217...73N}. 

Thus far, decayless oscillations have mostly been detected using extreme ultraviolet (EUV) imaging data taken with the Atmospheric Imaging Assembly (AIA), onboard the Solar Dynamics Observatory. The AIA images have a spatial resolution of about 1.4\arcsec.
In this work, we employ high resolution EUV imaging data from the Extreme Ultraviolet Imager \citep[EUI;][]{2020A&A...642A...8R} on board the recently launched Solar Orbiter \citep[][]{2020A&A...642A...1M} to re-investigate the phenomenon of decayless oscillations in active region coronal loops. At the time of our observations, Solar Orbiter was at a distance of $\approx$0.5\,AU, nearly along the Sun-Earth line. These data have a spatial resolution that is more than two times higher than that of the data from the AIA instrument (see Sect.\,\ref{sec:data} for details). This allows us to complement higher-resolution EUI observations with co-temporal EUV images from SDO/AIA to address the sub-arcsec nature of decayless oscillations for the first time. Here we present observations of decayless oscillations in a system of active region loops rooted in sunspots. Our study provides important insights into the drivers of decayless oscillations.

\begin{table*}[!ht]
\caption{Details of the datasets used in this study.}  
\label{table}
\begin{center}
\centering
   \begin{tabular} {@{}clccccc@{}}

         \hline

   Instrument & Channel & Time  & D$\mathrm{_{Sun}}$ & Angle with & Pixel & Cadence  \\
              &         & (UT)  & a.u & Sun-Earth line (deg) & scale (km) & (s)  \\

     \hline
       \rule{0pt}{2ex}\textbf{$\mathrm{Date : 2022-03-03}$} &  &   & \\     \hline
       \rule{0pt}{2ex}EUI & 174 & 09:43 - 10:43 & 0.54 & -6.0 & 195 & 5 \\
       AIA & 171, 193, 94  & 08:43 - 11:43 & 0.99 & - & 431 & 12  \\
       AIA & 1600  & "  & " & - & " & 24   \\
       HMI & LOS  & " & " & - & 359 & 45   \\

     \hline
       \rule{0pt}{2ex}\textbf{$\mathrm{Date : 2022-03-04}$} &  &   & \\     \hline
       \rule{0pt}{2ex}EUI & 174 & 10:48 - 11:48 & 0.53 & -4.6 & 190 & 5   \\
       AIA & 171, 193, 94  & 10:00 - 12:30 & 0.99 & - & 431 & 12   \\
       AIA & 1600  & " & " & - & " & 24  \\
       HMI & LOS  & " & " & - & 359 & 45  \\
       
    \hline
       \rule{0pt}{2ex}\textbf{$\mathrm{Date : 2022-03-05}$} &  &   & \\     \hline
       \rule{0pt}{2ex}EUI & 174 & 15:23 - 16:23 & 0.52 & --2.9 & 185 & 5  \\
       AIA & 171, 193, 94  & 15:00 - 17:00 & 0.99 & - & 431 & 12  \\
       AIA & 1600  & " & " & - & " & 24  \\
       HMI & LOS   & " & " & - & 359 & 45 \\
  \hline

\end{tabular}
\end{center}
\end{table*}

\section{Data\label{sec:data}}

In this study, we use extreme ultraviolet (EUV) imaging data covering the active region $\mathrm{NOAA~12957}$ between 2022-03-03 and 2022-03-05. These data were taken with the High Resolution Imager at 174~\AA\ (HRI$_{\rm{EUV}}$) of EUI\footnote{We use Level-2 EUI data available at \url{https://doi.org/10.24414/2qfw-tr95}} which followed this active region (AR) for three successive days starting from 2022-03-03, for about 1-hour on each day\footnote{There are EUI observations of the same active region post 2022-03-05 also. However, in those observations, either only a part of the active region was captured or the observations were recorded with a significantly lower cadence.}. At the time of these observations, Solar Orbiter was located close to the Sun-Earth line (see Table~\ref{table}) and hence, we complement EUI observations with data from the Solar Dynamics Observatory \cite[SDO;][]{2012SoPh..275....3P}. In particular, we use EUV/UV images from the Atmospheric Imaging Assembly \cite[AIA;][]{2012SoPh..275...17L} and line-of-sight (LOS) magnetograms from the Helioseismic and Magnetic Imager \cite[HMI;][]{2012SoPh..275..207S}, both onboard SDO. Details of each of the datasets used in this study are presented in Table~\ref{table}. The alignment between AIA and EUI images were done by combining information from the FITS headers and visual inspections. Lastly, we take into account the difference in light travel times between Sun-EUI and Sun-SDO while analysing co-temporal EUI-SDO datasets. In connection to this, all the time stamps quoted in the paper are Earth times.

\begin{figure}[!htb]
\centering
\includegraphics[width=0.49\textwidth,clip,trim=0.3cm 1cm 3.2cm 0cm]{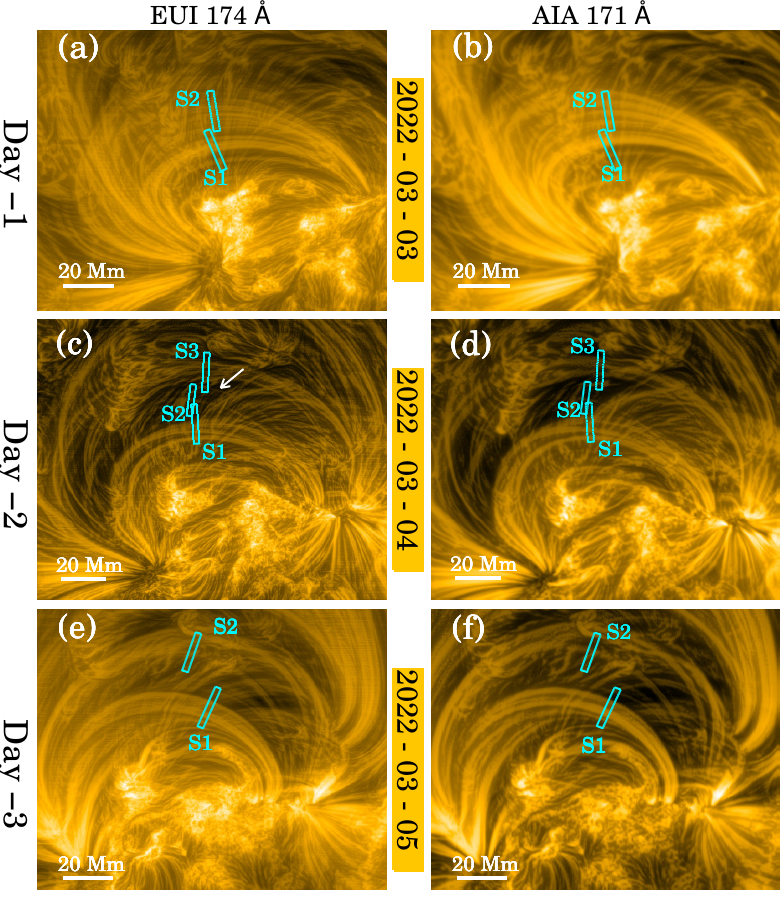}
\caption{Active region NOAA~12957. Overview of the observations from EUI (left column) and AIA (right column). In each panel, the thin blue boxes mark the positions of the artificial slits that are used for generating X-T maps. The white arrow in panel-c points to a particular thread that is only visible in EUI. Images shown in all panels are averaged over the corresponding observation time of 1-hour. An animated version of this figure is available \href{https://drive.google.com/file/d/1cCI7Gm4id20mDZPJct19VSNljuFE3QNF/view?usp=sharing}{here}.
}
\label{fig:context}
\end{figure}
\begin{figure}[!htb]
\centering
\includegraphics[width=0.49\textwidth,clip,trim=0cm 0cm 0cm 0cm]{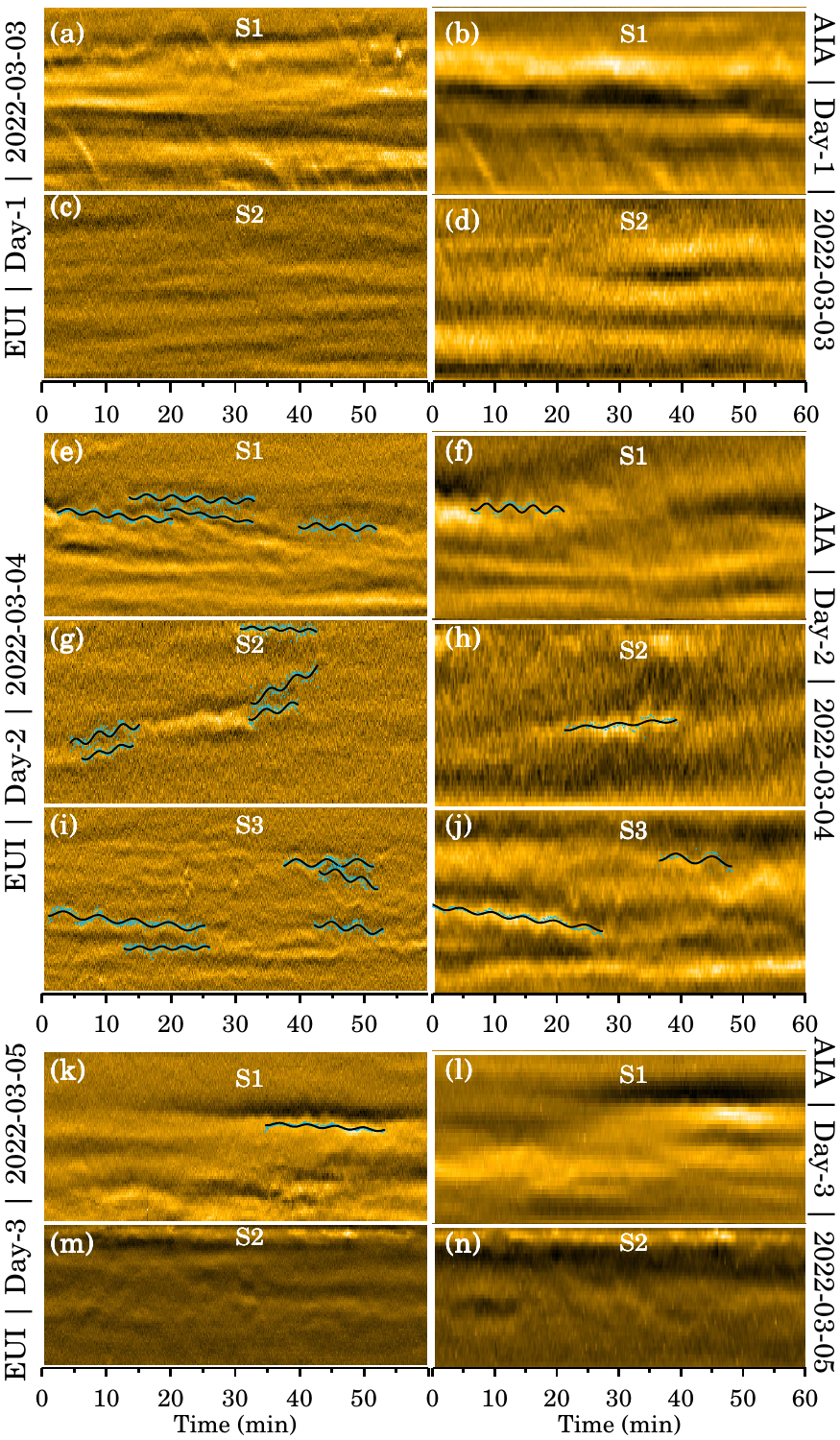}
\caption{Decayless transverse oscillations. Space-Time (X-T) maps derived from EUI (left column) and AIA (right column) image sequence. In some panels, the blue points highlight the individual oscillating features, wherein the black curves represent the best fit function (Eqn.~\ref{equation1}) to those. An animated version of this figure is available \href{https://drive.google.com/file/d/1WTWvwMeIG43CWCV5v_2UpAdUny7oQkR3/view?usp=sharing}{here}
}
\label{fig:xt}
\end{figure}

In order to remove effects of spacecraft jitter from EUI images, we used a cross-correlation based image alignment technique. This is done by first dividing the full image sequence into shorter, overlapping image sequences, in which, the last image from a given shorter sequence would be the same as the first image in the following shorter sequence. Then all the images in a given shorter sequence are co-aligned to the first image in that sequence. This would result in dataset in which all the images are aligned to the first image of the whole observing sequence considered.


\section{Results}\label{sect:results}

All three EUI observing sequences used in this study primarily imaged closed, nearly semi-circular (inferred visually from their plane of sky projections) coronal loops in $\mathrm{NOAA~12957}$. These loops as observed with EUI and AIA are displayed in Fig.\,\ref{fig:context}. Although images from the EUI 174\,{\AA} and the AIA 171\,{\AA} filter appear quite similar to each other,  differences, specially in terms of their multi-threaded structuring, are also visible. For example, the thread that we have highlighted with a white arrow on Day-2 EUI image (Fig.~\ref{fig:context}c) is not visible in the corresponding AIA 171~{\AA} image (Fig.~\ref{fig:context}d). This is not only due to the roughly two times better spatial resolution of EUI over AIA,  it is also related to the differences in passband response of HRI$_{EUV}$ and AIA 171~{\AA}. See Sec.~\ref{sect:appa} for a discussion on this. Lastly, we take a note on the plethora of background features that we find in Fig.~\ref{fig:context}. It includes moss type structures as well as short background loops (especially on Day-3; Fig.~\ref{fig:context}e-f). 
Nonetheless, looking at the event movie (available online), we notice prominent loop oscillations on Day-2 image sequence whereas oscillations (if any) on other two days are not so obviously detectable in the movie. Thus, our aim is to first quantitatively analyse these oscillations and then search for the source that drives them. 


\begin{figure}[!htb]
\centering
\includegraphics[width=0.45\textwidth,clip,trim=0cm 0cm 0cm 0cm]{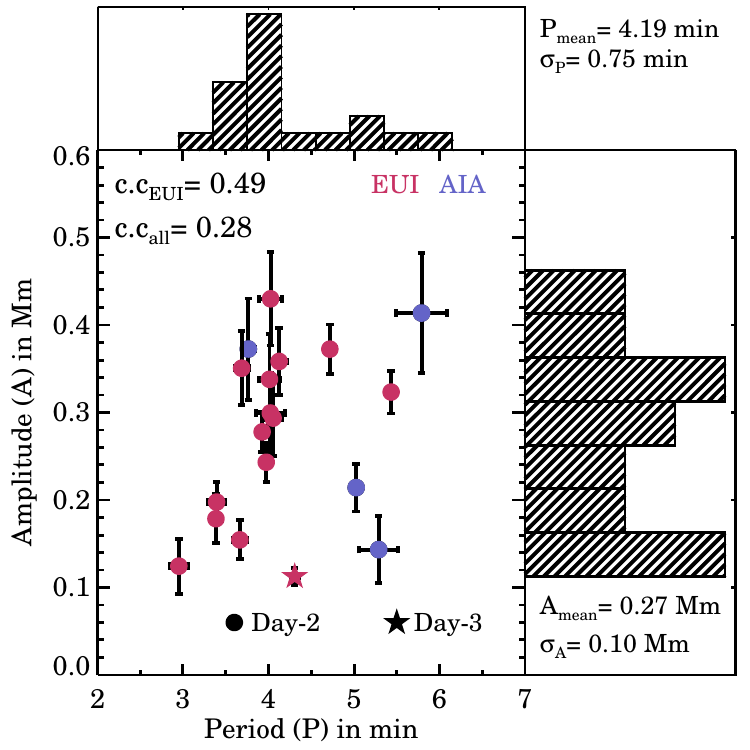}
\caption{Period-Amplitude scatter plot. Symbols (circle and star) highlight the two observation days wherein the two colors magenta and blue represent observations from EUI and AIA, respectively. Period vs. amplitude cross-correlation (c.c) values (one with all data points and another with only EUI data) are indicated.  The top and side panels show the period and amplitude histograms.}
\label{fig:pa_scatter}
\end{figure}
\begin{figure*}[!htb]
\centering
\includegraphics[width=0.90\textwidth,clip,trim=0cm 0cm 1.5cm 0cm]{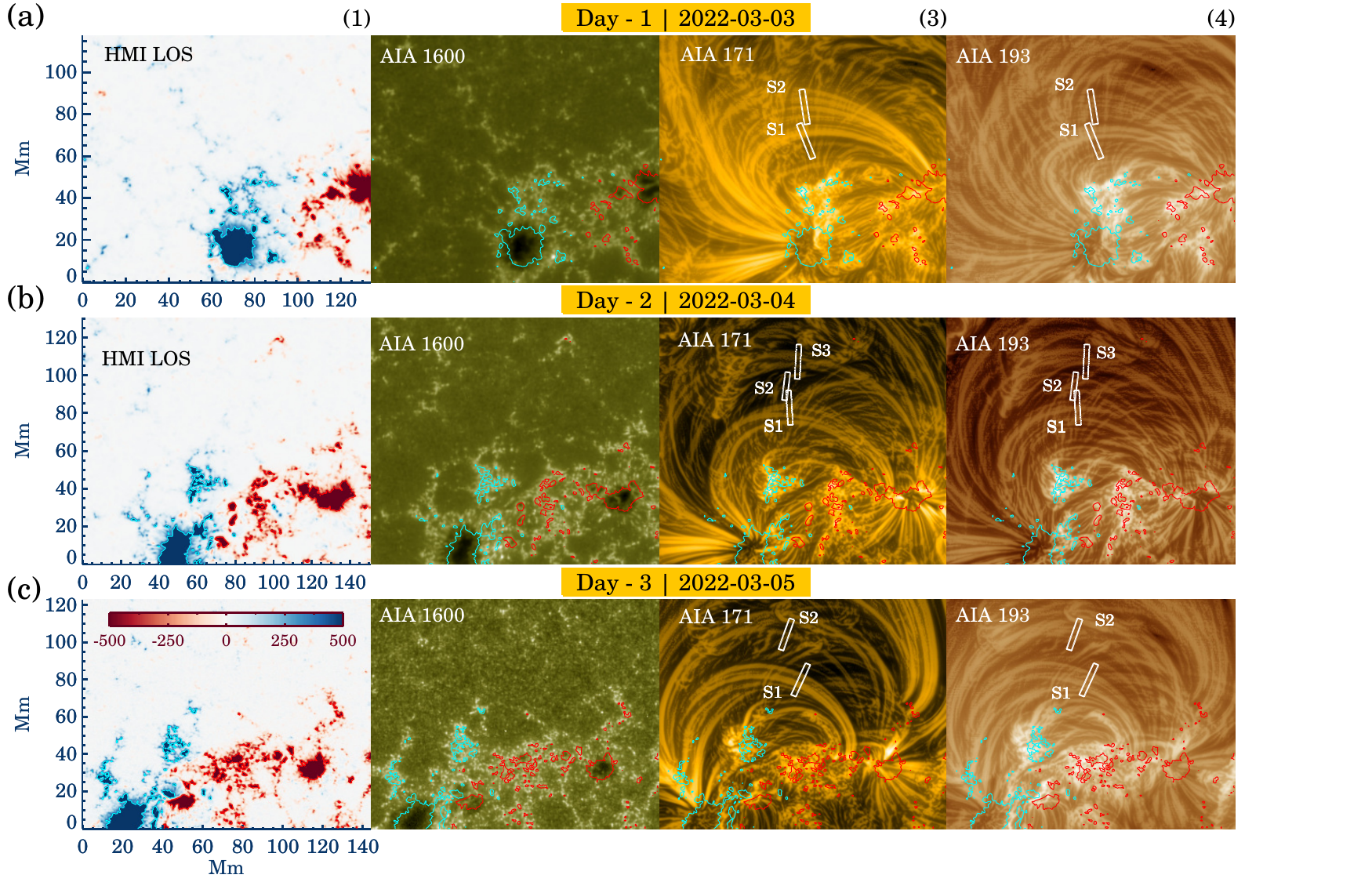}
\caption{Magnetic configuration of the loop footpoints. Along reach row, the first panel shows the time average LOS magnetogram from HMI, followed by time average images from AIA 1600~{\AA}, 171~{\AA} and, 193~{\AA} channels. The blue and red contours overplotted on top of every image of a given row represent boundaries of $\pm$400 G as derived from the time averaged HMI magnetogram of that day. }
\label{fig:magnetic}
\end{figure*}
\begin{figure}[!htb]
\centering
\includegraphics[width=0.49\textwidth,clip,trim=0cm 0cm 0cm 0cm]{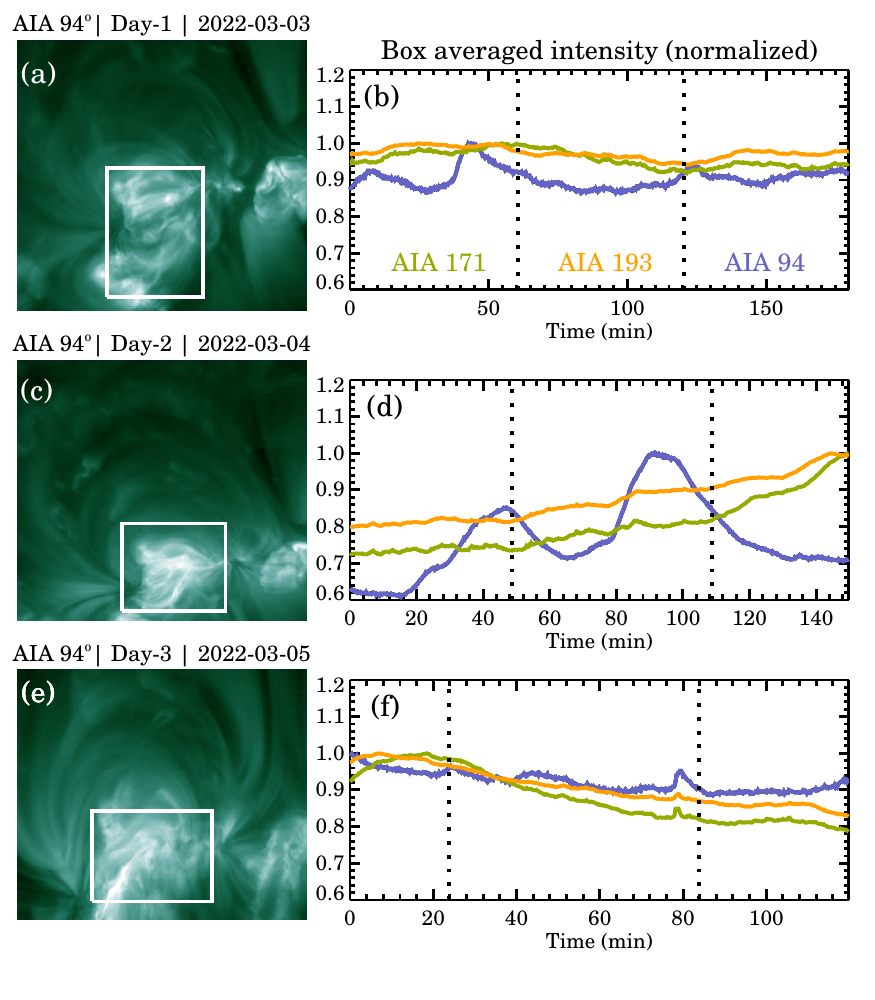}
\caption{Analysis of AIA Light curves. Left panels (a,c,e) show the 94~{\AA} images of corresponding days along with the locations (white rectangular boxes) from where the light curves (panels-b,d,f) are derived. The vertical dotted lines in panels-b,d,f mark the start and end of the EUI observing periods.}
\label{fig:light}
\end{figure}
\subsection{Capturing transverse oscillations}\label{sect:xt}

In order to capture the observed transverse oscillations, we created space-time (X-T) maps by placing artificial slits perpendicular to the length of the loops. Final slit locations (the blue rectangular boxes in Fig.~\ref{fig:context}) were selected by first placing multiple such slits covering the entire length of a given loop and then picking out only those slits along which we find (visually) prominent oscillation signals\footnote{None of the Day-1 X-T maps show any oscillations and hence, the slits shown in Fig.~\ref{fig:context}a,b are the ones for which the maps are relatively free from any dynamic background features.}. All panels of Fig.~\ref{fig:xt} present the  X-T maps derived for the given slit in Fig.~\ref{fig:context}. Moreover, to further enhance the appearance of the oscillating threads in these X-T maps, a smooth version of the map (boxcar-smoothed in the transverse direction) is subtracted from each original map.
Prominent oscillation signatures are found on Day-2 along all the three slits, both in the EUI and AIA data (Fig.~\ref{fig:xt}e-j). Interestingly on Day~3, we find such oscillatory pattern in only one of the EUI maps (derived from slit-$\rm{S1}$; Fig.~\ref{fig:xt}k) whereas the corresponding co-temporal AIA map shows little oscillatory signal. On the contrary to the maps from these two days, none of the Day-1 maps show any transverse oscillation signal. This suggests that coronal loops within the same magnetic system exhibit decayless oscillations somewhat sporadically.  Furthermore, by looking at the EUI and AIA images in Fig.~\ref{fig:context}, we find that  significantly more  threads are present on Day-1 images as compared to Day-2 and Day-3. The presence of such internal fine structure within a loop bundle could lead to overdamping of waves, as hinted in studies by \citet{2010ApJ...716.1371L} and \citet{2014A&A...570A..84N}, and could potentially be an explanation of why the oscillations are absent in Day-1 datasets.

Before we delve further into this question of why these oscillation signals appear so pronounced on a particular day but are mostly absent on other days, let us first take a closer look at the oscillations themselves. At a first glance, it is evident that the oscillations that we find on Day-2 and Day-3, do not show any considerable decay in their amplitudes over multiple wave periods\footnote{There are some oscillations (e.g. in Fig.~\ref{fig:xt}g) which show no decay in their amplitudes but last only for two wave periods. We still consider them in our analysis since most of those oscillations appear on threads that exhibit such decayless behaviour multiple times during the observation.}. Hence, these waves clearly belong to the category of `decayless' oscillations \citep{2013A&A...560A.107A}.
In order to quantitatively analyse these oscillations, we first identify the location of an individual strand in the X-T map at each time step by fitting a Gaussian along the transverse direction of that strand. We then fit these derived positions of the strand as a function of time following the function:
\begin{equation}
\label{equation1}
\centering
 y(t)=A \left (\sin{\dfrac{2\pi t}{P}+\phi} \right)+c_1 t+c_0,
\end{equation}
where $A$ is the oscillation amplitude, $P$ represents the period, $\phi$ is the phase, and  $c_{0}$ and $c_{1}$ are constants. Results based on these fits are shown in Fig.~\ref{fig:xt}.  The blue points highlight the detected Gaussian mean of a given strand at each time-step, while the black curves are the corresponding fitted function.

We then study the relation between the derived periods and amplitudes of these waves. Figure~\ref{fig:pa_scatter} shows the scatter plot between the fitted $A$ and $P$ values. As seen from this plot, the detected periods lie between 2 and 7 min ($P_{\rm avg}=4.19$ min, $\sigma=0.75$ min) while the displacement amplitudes are in the range of 0.1 to 0.5 Mm ($A_{\rm avg}=0.27$ Mm, $\sigma=0.10$ Mm). These numbers are quite comparable to the values obtained by \citet{2015A&A...583A.136A}, who performed a statistical study of these waves using data from AIA. Additionally, from Fig.~\ref{fig:pa_scatter}, we also find a weak but positive correlation between oscillation periods and amplitudes. The Pearson correlation is estimated to be 0.28 (with the null hypothesis probability $\rm{p}$=0.24) when considering all data points, while it is 0.49 (with $\rm{p}$=0.07) with only EUI data. Hence, from our sample set, we conclude that this is not a statistically significant relationship. However, using a much larger sample set from AIA, \citet{2022MNRAS.tmp.1009Z} also reported the presence of such a positive correlation between these two quantities.


\subsection{Magnetic configuration of the loop footpoints}
 
We shift our focus now towards finding the origin of these oscillations. One of the ways to explain the observed decayless oscillations is through self-oscillation. As mentioned in the introduction, \citet{2016A&A...591L...5N} proposed a scenario in which the footpoints of the oscillating loops get perturbed by supergranular flows which then lead the system into a self-oscillation state. To check whether such a mechanism is in operation also in our case, we first investigate whether there is any change in the footpoint configuration on Day-2 (when these oscillations were widely present) as compared to Day-1 (when no oscillations were seen). 

To this end, we concentrate on the magnetic configuration of this loop system using LOS magnetic field data from HMI. In different panels of Fig.~\ref{fig:magnetic}, we show the HMI data for the same FOV as in Fig.~\ref{fig:context} for all three days. Alongside, we also show images from AIA 1600\,\AA, 171~{\AA}, and 193~{\AA} channels. Looking at the Day-1 HMI magnetogram (panel \ref{fig:magnetic}a-1), we find that the AR has a clear bipolar structure with opposite polarities being well separated from each other. The same structure through the 1600~{\AA} channel (which mainly captures the uppermost photosphere) reveals two well developed sunspots along with a pore near the western spot. To locate photospheric footpoints of the coronal loops under investigation, we overplot the contours of $\pm$400~G as derived from the HMI data on every other panel. Through this we find that for most loops, both footpoints are rooted within sunspots (either inside or at the edge of umbrae; Fig.~\ref{fig:pa_scatter}a-3,a-4). 
The situation remains more-or-less similar on the other two days. HMI images of these two days (panel \ref{fig:magnetic}b-1,c-1) reveal that the two dominant polarities of the active regions have further evolved and separated compared to the scene on Day-1. Signatures of new flux emergence are also noted in these magnetograms. Similar to before, by following the HMI contours on 171~{\AA} and 193~{\AA} images, we find the west side footpoint of one of the oscillating loops (slit-$\rm{S1}$) on Day-2 is now (partly) rooted in a pore like structure. 

In summary, we find that the oscillating loops are mostly rooted in regions of strong magnetic fields and the whole configuration does not show any significant change over three days of this observation. Moreover, it is known that sunspots or pores are devoid of any supergranular motions and hence the scenario of footpoint driving that we mentioned earlier seems not to be the case here.  However, it remains to be further addressed whether horizontal motions, such as those observed in penumbral filaments or the large-scale flows that drive the sunspots apart, could also lead to these decayless oscillations.

\subsection{Activity near the loop footpoints }

While looking at the event movie (available online), we noticed that although 171~{\AA} and 193~{\AA} filters do not show much coronal (flaring) activity, the 94~{\AA} movie (which also samples flaring plasma of 5--7~{\rm MK} temperature) seems to show quite a bit of intensity enhancements and brightenings in-between two loop footpoints. These intensity enhancements are related to low level flaring activity in the AR core \citep[see e.g.][]{2020A&A...644A.130C}. To explore whether these small flares play any role in generating decayless oscillations, we analyse AIA light curves extracted from a large enough region encompassing the brightening AR core as shown by the white boxes in Panels~\ref{fig:light}a,c,e. Panels~\ref{fig:light}b,d,f present the extracted curves (normalised to their maximum values). Looking at the Day-2 light curves, we find two clear peaks in the 94~{\AA} data whereas the 171~{\AA} and 193~{\AA} curves do not show similar features (these curves show a gradual increasing trend). 

Thus, at this point, we are tempted to conclude that the prominent oscillation signals that we find in Fig.~\ref{fig:xt}e,g,i are causally connected with these intensity enhancements. However, light curves from the two other days do not exactly fit into this scheme. For example, on Day-1, we also see a peak in the 94~{\AA} channel, just before the EUI observation began on that day. However, the corresponding X-T maps (Fig.~\ref{fig:xt}a,c) do not contain any oscillation signal. On the contrary, on the Day-3 X-T map (Fig.~\ref{fig:xt}k) we find an oscillation signal without any associated intensity enhancement in any of the AIA channels (the small peak near t$\approx$80 min appears much later than the oscillations in the X-T map). 
Hence, it is inconclusive at this moment whether there is any causal link between the core brightenings that we observe in 94~{\AA} channel and the appearance of the decayless oscillations in any of the loops. Furthermore, the brightenings in the 94~{\AA} channel appear to be spatially disconnected from the oscillating loops or their footpoints. Based on these evidences, we suggest that the observed flaring activity is not very likely to contribute towards the observed decayless oscillations.

\begin{figure}[!htb]
\centering
\includegraphics[width=0.50\textwidth,clip,trim=0cm 0cm 0cm 0cm]{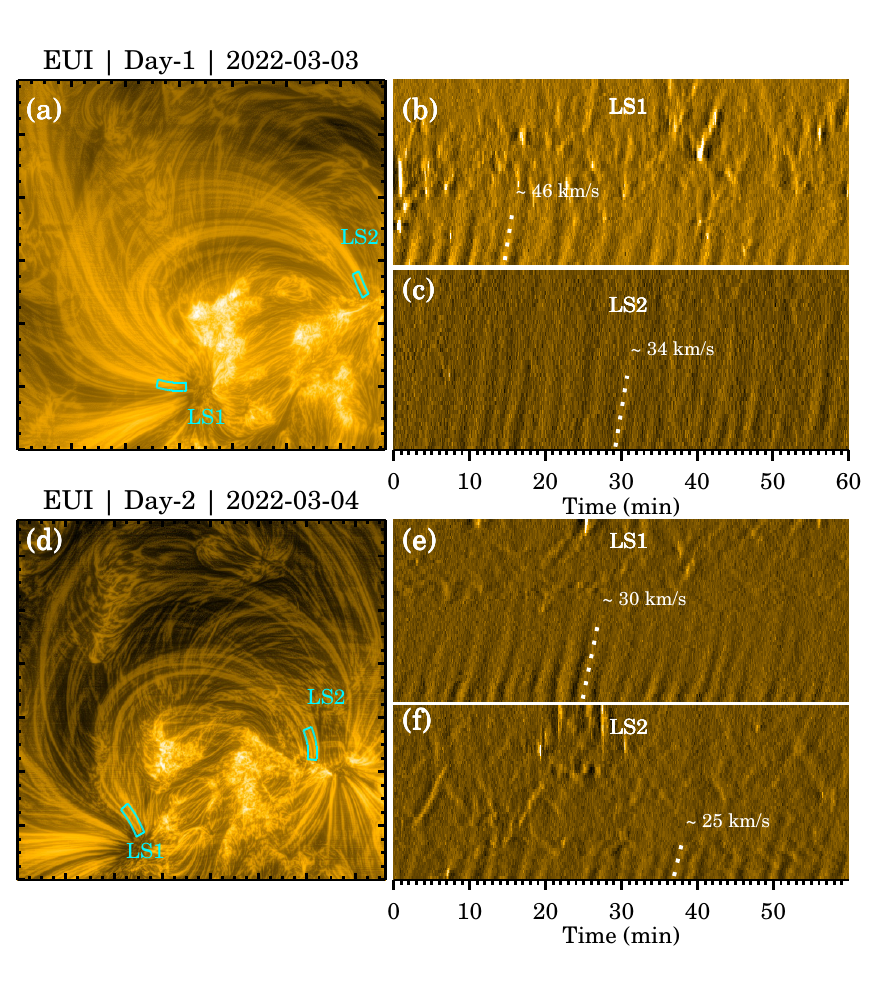}
\caption{Propagating intensity disturbances at the loop footpoints. Green boxes in Panels-a,d show the locations of the artificial slits that we used to generate the X-T maps shown in Panels-b,c,e,f. The slanted dotted white lines in these panels are representative of slopes of the individual ridges that we used for speed estimation.}
\label{fig:longitudinal}
\end{figure}
\subsection{Longitudinal perturbations at the footpoints}

Finally, we explore the possibility of longitudinal perturbations (near the loop footpoints) being the driver of the decayless oscillations. It is commonly observed that footpoints of AR loops at coronal height host repetitive propagating disturbances whose periods lie between 3 to 5 mins \citep{2000A&A...355L..23D}. These features are generally explained in terms of p-modes leakage from the lower atmosphere \citep{2005ApJ...624L..61D}. There have been analytical as well as numerical studies in the past (e.g., \citealp{2019A&A...625A.144R}) which show that these p-modes can in principle undergo mode conversion and manifest themselves as transverse oscillations. Again, if this mechanism is in operation, then we would expect a difference (due to some unknown reason) in these longitudinal perturbations on Day-1 (when oscillations are not detected) as compared to Day-2 (when prominent oscillations are seen). To test this hypothesis, we generate X-T maps by placing slits near the footpoints of those loops that show transverse oscillations. The green rectangular boxes shown in Fig.~\ref{fig:longitudinal}a,d outline these slits wherein the derived maps are shown in Fig.~\ref{fig:longitudinal}b,c,e,f. On both days, these maps show clear signatures of repetitive propagating intensity disturbances at sub-sonic speeds (approximately 30 km~s$^{-1}$). Thus, we conclude that there is nothing special with these longitudinal perturbations at the footpoints of coronal loops on Day-2, which could give rise to the observed decayless oscillations.

\section{Summary and Discussion}
In this letter, we revisited the phenomena of decayless transverse oscillations in AR loops. Using high resolution, high cadence EUV imaging data, we followed a system of AR loops for three consecutive days while it was passing the solar disc. Below we highlight our main findings.

\begin{itemize}

\item One of our key results is related to the better visibility or detectability of these oscillations in high resolution EUI data (panels~\ref{fig:xt}e,g,i) in comparison to AIA (panels~\ref{fig:xt}f,h,j). We conjecture that it is not solely due to the spatial resolution difference between these two instruments, but rather has contributions from the differences in their filter response as well as instrument sensitivity. Nonetheless, if our finding also holds for other loops, then it has a profound impact on the calculation of total energy that is carried by these waves as we can now resolve significantly more of these on any given loop, at any given time.\\

\item Interestingly, the same system of loops does not always show decayless oscillations, although the physical conditions remain more-or-less the same.\\

\item By studying the magnetic configuration of the loop footpoints (Fig.~\ref{fig:magnetic}), we found that some of these loops were rooted in sunspots which are known to be devoid of supergranular flows. Such flows were previously considered to be one of the ways to drive the loop footpoints and in turn, generate these decayless oscillations. However, our analysis shows that it is not the case in this event and hence, the search for that elusive wave driver needs to continue.\\

\item We have explored the possibility of a localised heating event being the driver of these oscillations (Fig.~\ref{fig:light}). We found no clear evidence for a causal connection between localised heating events and the appearance of these oscillations. The same also applies for the p-mode scenario (Fig.~\ref{fig:longitudinal}). Moreover, we found no clear changes in the behaviour of longitudinal waves at the loop footpoints driven by p-modes over the time of observations, while the decayless oscillations were clearly more sporadic.

\end{itemize}

 In conclusion, through our analysis we found that none of the commonly suspected sources proposed to drive decayless oscillations in active region loops (e.g., supergranular flows, transient events and mode conversion near the loop footpoints) seemed to be operating in the studied events. This however, does not rule out the possibility that one of the above mentioned sources may still be the dominant driver of decayless oscillations seen in loops that are not rooted in sunspots. Further studies are necessary to address the nature of decayless oscillations in loops rooted in different magnetic environments in the photosphere. To this end, EUV data with even higher resolution e.g., from EUI, during its closest perihelion passage, would be best suited.

\begin{acknowledgements}
We thank the anonymous reviewer for the encouraging comments and helpful suggestions. L.P.C. gratefully acknowledges funding by the European Union. Views and opinions expressed are however those of the author(s) only and do not necessarily reflect those of the European Union or the European Research Council (grant agreement No 101039844). Neither the European Union nor the granting authority can be held responsible for them. S.P. acknowledges the funding by CNES through the MEDOC data and operations center. D.M.L. is grateful to the Science Technology and Facilities Council for the award of an Ernest Rutherford Fellowship (ST/R003246/1). P.A. acknowledges STFC support from Ernest Rutherford Fellowship grant number ST/R004285/2. The ROB team thanks the Belgian Federal Science Policy Office (BELSPO) for the provision of financial support in the framework of the PRODEX Programme of the European Space Agency (ESA) under contract numbers 4000134474 and 4000136424. Solar Orbiter is a space mission of international collaboration between ESA and NASA, operated by ESA. The EUI instrument was built by CSL, IAS, MPS, MSSL/UCL, PMOD/WRC, ROB, LCF/IO with funding from the Belgian Federal Science Policy Office (BELSPO/PRODEX PEA 4000134088); the Centre National d’Etudes Spatiales (CNES); the UK Space Agency (UKSA); the Bundesministerium f\"{u}r Wirtschaft und Energie (BMWi) through the Deutsches Zentrum f\"{u}r Luft- und Raumfahrt (DLR); and the Swiss Space Office (SSO). AIA is an instrument on board the Solar Dynamics Observatory, a mission for NASA's Living With a Star program. AIA and HMI data are courtesy of NASA/SDO and the AIA, EVE, and HMI science teams. This research has made use of NASA’s Astrophysics Data System. The authors would also like to acknowledge the Joint Science Operations Center (JSOC) for providing the AIA data download links.
\end{acknowledgements}

\bibliographystyle{aa}
\bibliography{ref_decayless}


\clearpage

\begin{appendix}

\section{Comparison of loop visibility between EUI and AIA data}\label{sect:appa}

\begin{figure}[!htb]
\centering
\includegraphics[width=0.49\textwidth,clip,trim=0cm 0cm 0cm 0cm]{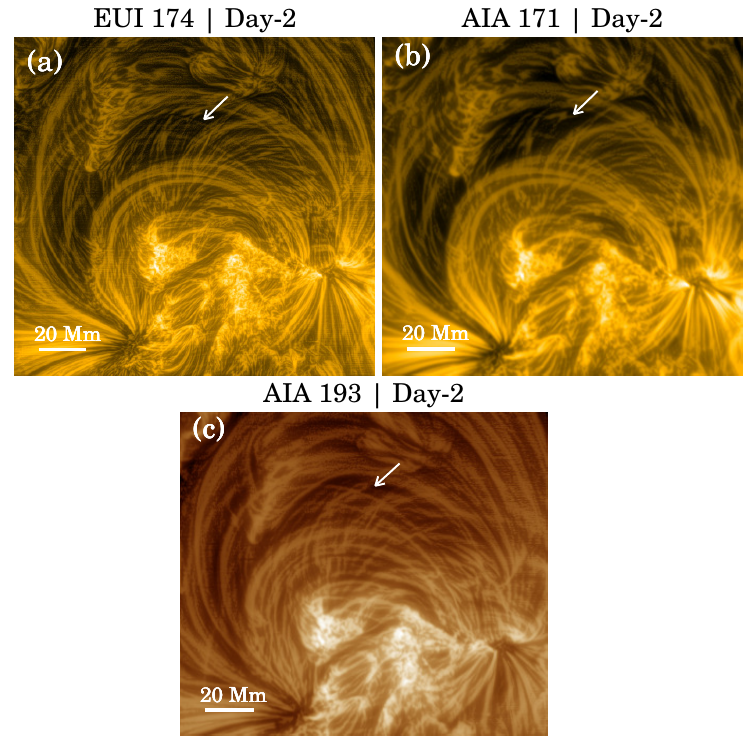}
\caption{Comparison of EUI and AIA images. Arrows point towards a particular loop strand that we identified in the EUI image. See text for details. }
\label{fig:193_171_comp}
\end{figure}

As mentioned in Sec.~\ref{sect:results}, certain features that can be identified in EUI 174~{\AA} images, are missing in AIA 171~{\AA} channel data but visible in AIA 193~{\AA} data. The loop thread that we highlight with arrows in Fig.~\ref{fig:193_171_comp} (and also in Fig.~\ref{fig:context}c) is an example of that. From these evidences we conclude the following: 1) the thread is a multi-thermal structure as it appears simultaneously in EUI 174~{\AA} and AIA 193~{\AA}. However, the scenario in which the thread is an isothermal structure with a temperature that falls between the peak response of EUI 174~{\AA} and AIA 193~{\AA}, is also a possibility, and 2) due to the slight shift in the peak of EUI 174~{\AA} bandpass towards somewhat higher temperatures than AIA 171{\AA} channel (see Fig. 1 of \citealp{2021A&A...656L...7C}), this loop only appears in EUI but not in AIA~171{\AA}. 

\end{appendix}

\end{document}